\documentclass{PoS}

\title{A MEERKAT VIEW ON GALAXY CLUSTERS}

\ShortTitle{A MeerKAT view on galaxy clusters}

\author{\speaker{G.~Bernardi}$^{1,2}$, T.~Venturi$^3$, R.~Cassano$^3$, G.~Brunetti$^3$, D.~Dallacasa$^4$, B.~Fanaroff$^1$, B.~Hugo$^1$, S.~Makhatini$^1$, N.~Oozeer$^{1,5,6}$, O.M.~Smirnov$^{2,1}$ and J.T.L.~Zwart$^{7,8}$\\
	$^1$ SKA SA, 3rd Floor, The Park, Park Road, Pinelands, 7405, South Africa. E-mail: \email{gbernardi@ska.ac.za, bfanaroff@ska.ac.za, bhugo@ska.ac.za, smakhathini@ska.ac.za, nadeem@ska.ac.za, oms@ska.ac.za};\\ 
	$^2$ Department of Physics and Electronics, Rhodes University, PO Box 94, Grahamstown, 6140. E-mail: \email{g.bernardi@ru.ac.za, o.smirnov@ru.ac.za};\\
	$^3$ INAF-IRA, via Gobetti 101, 40129 Bologna, Italy. E-mail: \email{tventuri@ira.inaf.it, r.cassano@ira.inaf.it, brunetti@ira.inaf.it};\\
	$^4$ Dipartimento di Fisica e Astronomia, Universit\'a di Bologna, viale Berti Pichat 6/2, I-40127 Bologna, Italy. E-mail: \email{daniele.dallacasa@unibo.it};\\
	$^5$ South Africa; African Institute for Mathematical Sciences, 6-8 Melrose Road, Muizenberg 7945, South Africa;\\
	$^6$ Centre for Space Research, North-West University, Potchefstroom 2520, South Africa;\\
	$^7$ Department of Physics \& Astronomy, University of the Western Cape, Private Bag X17, Bellville, Cape Town 7535, South Africa. E-mail: \email{jz@uwcastro.org};\\
	$^8$ Astrophysics, Cosmology \& Gravity Centre, Department of Astronomy, University of Cape Town, Private Bag X3, Rondebosch 7701, South Africa.} 

\abstract{Almost two decades of observations of radio emission in galaxy clusters have proven the existence of relativistic particles and magnetic fields that generate extended synchrotron emission in the form of radio halos. In the current scenario, radio halos are generated through re--acceleration of relativistic electrons by turbulence generated by cluster mergers. Although this theoretical framework has received increasingly supporting observational evidence over the last ten years, observations of statistically complete samples are needed in order to fundamentally test model predictions. In this paper we briefly review our 7--element Karoo Radio Telescope observations of a sample of nearby clusters aimed to test the predictions of the turbulent re--acceleration model in small systems ($M_{500} > 4 \times 10^{14}$~M$_{\odot}$). We conclude by presenting two galaxy cluster surveys to be carried out with MeerKAT in order to provide crucial test of models of radio halo formation in nearby ($z < 0.1$) and high redshift ($z > 0.4$) systems respectively.}

\FullConference{MeerKAT Science: On the Pathway to the SKA,\\
	25-27 May, 2016,\\
		Stellenbosch, South Africa}

\begin{document}

\section{Background and open questions}
Galaxy clusters are the largest gravitationally bound systems and are believed to be formed via mergers of smaller systems. They have masses of the order of $10^{14}-10^{15}$~M$_{\odot}$, with $15-20\%$ in the form of a hot ($10^8$~K) gas that pervades the cluster volume, emitting X--rays via the Bremsstrahlung mechanism and mm-wave radiation via the Sunyaev--Zeldovich (SZ) effect. The presence of a non--thermal (i.e. relativistic particles and magnetic fields) component emitting synchrotron radiation has been revealed by a variety of radio observations over the last few decades. In particular, there are four different sources of radio emission found in galaxy clusters (see \cite{feretti12} for a recent observational review):
\begin{itemize}
\item discrete radio sources associated with cluster galaxies;
\item radio halos (RHs): Mpc--scale diffuse radio sources with steep spectrum and low surface brightness that are found in the central regions of a number of merging clusters;
\item mini halos: central, diffuse radio sources extending over $\sim 100$~kpc scales, typical of dynamically relaxed systems;
\item radio relics: diffuse radio sources with elongated morphology, significantly polarized, mostly located at the outskirts of a small number of merging clusters.
\end{itemize}
In this paper we will mainly focus on the study of RHs and what they can tell us about the formation and evolution of galaxy clusters.

The particle life time due to radiative losses is much shorter than the RH crossing time (e.g. \cite{jaffe97}), therefore a re--acceleration mechanism is required to explain the presence of Mpc--size structures in galaxy clusters. Theoretical efforts over the last two decades have provided a scenario for the formation of RHs based on the {\it in--situ} re--acceleration of relativistic particles due to merger--driven turbulence (e.g., \cite{brunetti14} for a review). Radio observations of large samples of clusters carried out over the last ten years in the Extended GMRT Radio Halo Survey (EGRHS, \cite{venturi07}, \cite{venturi08}, \cite{kale13} and \cite{kale15}) have established two key properties:
\begin{itemize}
\item RHs are not ubiquitous, but are found only in the $20-30\%$ of clusters that are X--ray luminous (e.g., \cite{brunetti07}, \cite{cassano08});
\item RHs are connected to the cluster dynamical state, being found only in merging clusters but not in relaxed systems (\cite{cassano10}). This behaviour appears as a {\it bimodal} distribution between the 1.4~GHz RH power $P_{1.4}$ and the X--ray luminosity $L_{\rm X}$, where clusters with a RH follow the $P_{1.4} - L_{\rm X}$ correlation and clusters without a RH are significantly below such correlation (e.g. \cite{brunetti09});
\end{itemize}
With the advent of SZ cluster surveys (e.g., \cite{marriage11}, \cite{reichardt13}, \cite{planck14}), RHs could be directly linked to the cluster mass and their bimodality has been confirmed in the $P_{1.4} - M_{500}$\footnote{where $M_{500}$ is the total cluster mass within the radius $R_{500}$, defined as the radius corresponding to a total density contrast $500\rho_c(z)$, where $\rho_c(z)$ is the critical density of the Universe at  the cluster redshift.} plane too, with RH clusters following the correlation, whereas clusters without RHs appear well below the correlation (\cite{basu12}, \cite{cassano13}). In addition, it has been proven that the fraction of clusters with RH is larger in surveys of mass--selected clusters, being $\sim 50\%$ for clusters with $M_{500} > 6\times 10^{14} M_{\odot}$ (\cite{cuciti15}).

These observational results support the current scenario for the formation of giant RHs whose formation history depends on the interplay between the galaxy cluster merging rate throughout the cosmic epochs and the process of particle acceleration.

Despite the success of the turbulent re--acceleration model in explaining current RH observations, several questions related to the formation and evolution of diffuse radio emission in galaxy clusters are still open, in particular:
\begin{enumerate}
\item what is the fraction of giant radio halos in less massive clusters?
\item how does the fraction of clusters with giant radio halos evolve with cosmic epoch?
\end{enumerate}

\section{KAT--7 observations of galaxy clusters}
We recently began to address the first open question by selecting a mass-limited ($M_{\rm 500} > 4 \times 10^{14}$~M$_{\odot}$) sample of nearby ($z < 0.1$) clusters from the Planck SZ cluster catalogue (\cite{planck14}) and observed it with the 7--element Karoo Array Telescope (KAT--7, \cite{foley16}) at 1.86~GHz. The sample, consisting of 15 clusters observed with a $2.3 - 2.9$~arcmin angular resolution (depending upon declination), is described in \cite{bernardi16}, to which we refer the reader for details. We will summarize here the main results presented in the light of the future opportunities offered by the upcoming MeerKAT telescope described in Section~\ref{sec:MeerKAT_opportunities}.

\begin{figure*}
\centering
\includegraphics[width=0.49\columnwidth]{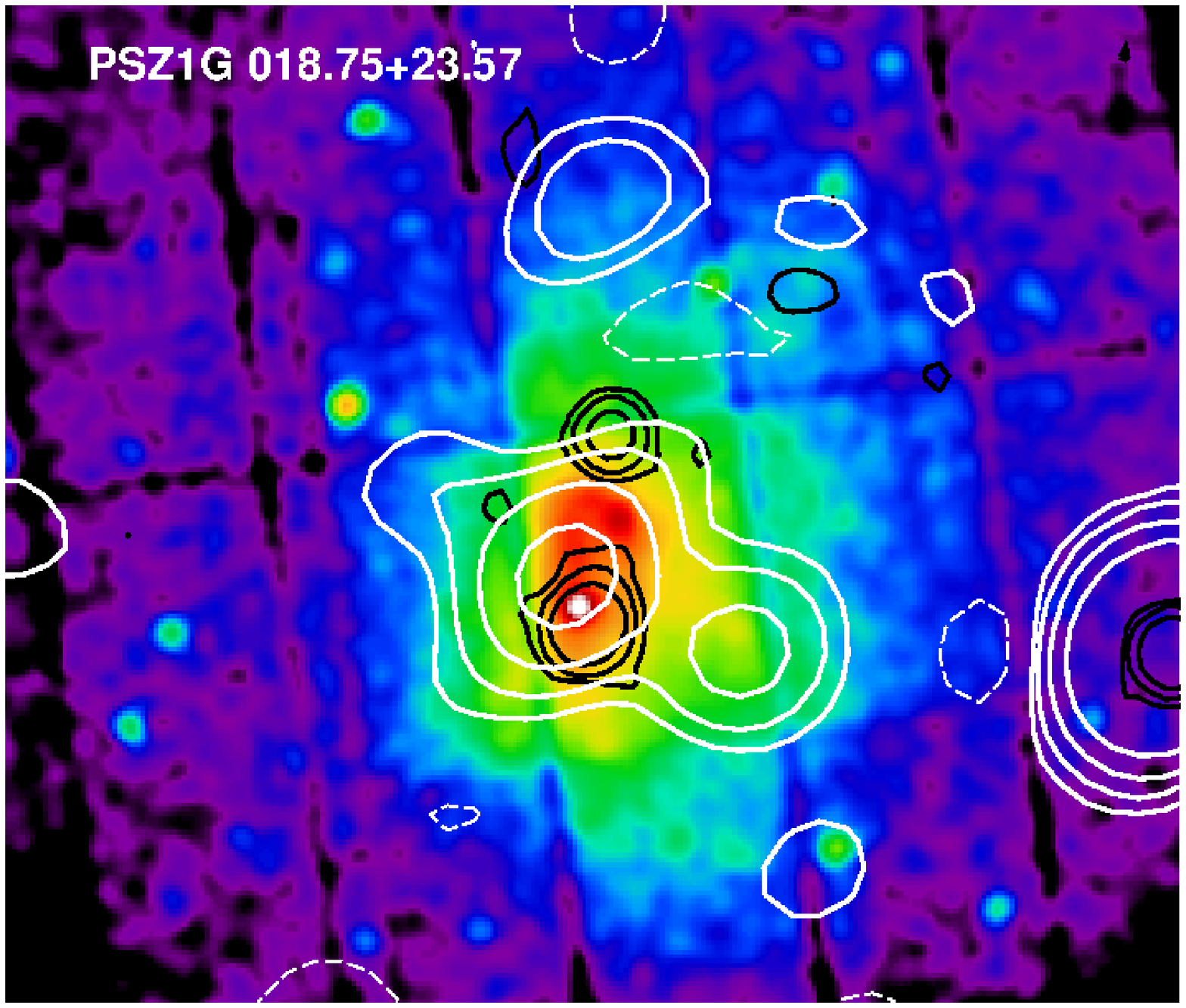}
\includegraphics[width=0.49\columnwidth]{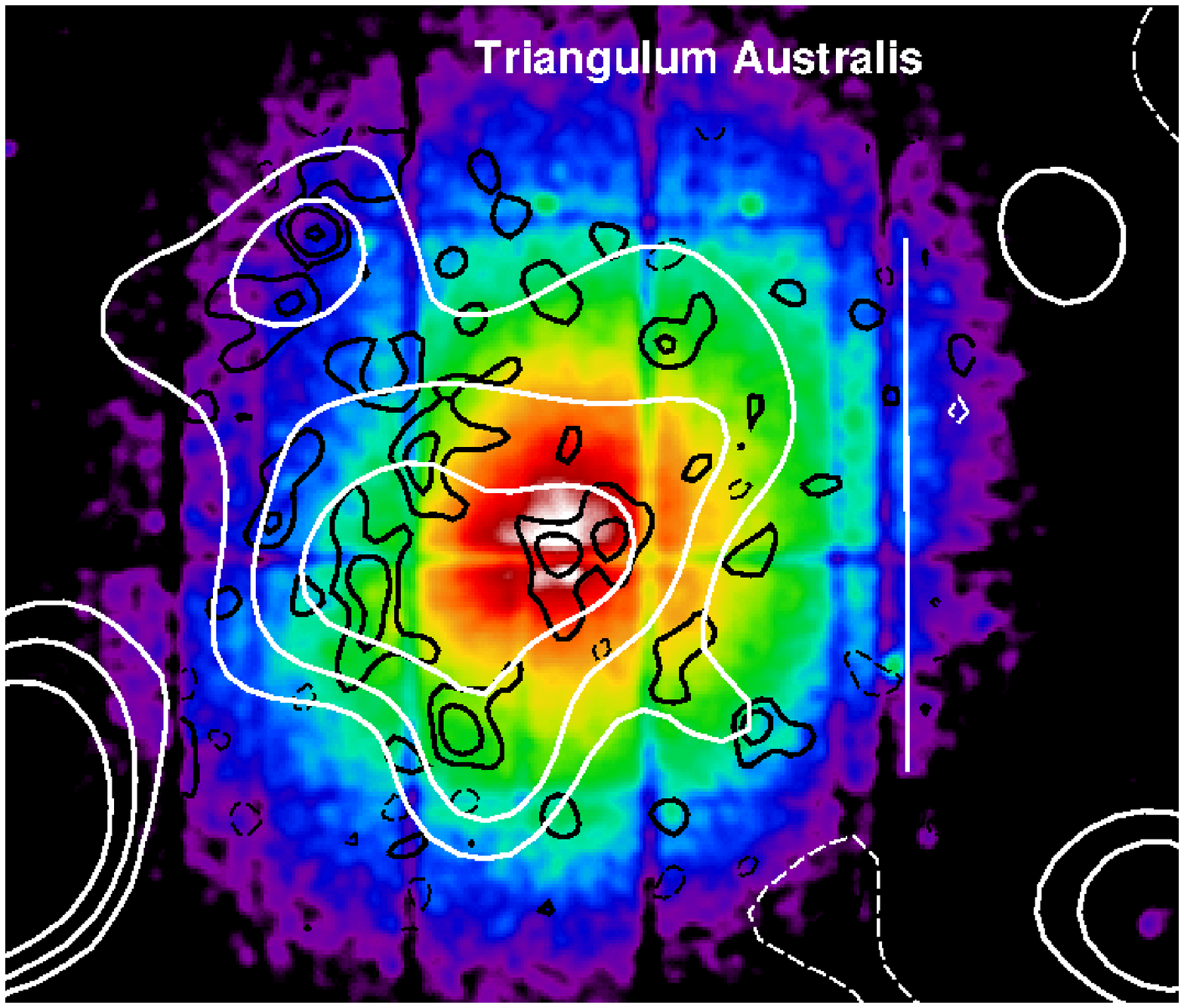}
\caption{1.86~GHz KAT--7 radio contours (white) overlaid on the {\it XMM--Newton} images for the PSZ1G\,018.75+23.57 (left) and Triangulum Australis (right) clusters as examples of candidate RHs (from \cite{bernardi16}). For PSZ1G\,018.75+23.57, contours are drawn at $\pm$~2.5, 5, 10, 20~mJy~beam$^{-1}$ (with a $2.8 \times 2.4$~arcmin synthesized beam) after subtracting the two NVSS compact sources (black contours, drawn at $\pm$~1.5, 6, 24~mJy~beam$^{-1}$). For Triangulum Australis, contours are drawn at $\pm$~2.5, 5, 10~mJy~beam$^{-1}$ (with a $2.9 \times 2.7$~arcmin synthesized beam) and SUMSS contours (black) are drawn at $\pm$~2, 4, 8~mJy~beam$^{-1}$. The vertical white bar indicates a 800~kpc size.}
\label{fig:candidates}
\end{figure*}

The results of KAT--7 observations can be divided in three groups: 
\begin{itemize}
\item six clusters which host radio emission. For these systems, the limited angular resolution prevented us from reliably disentangling compact sources blended in a possible diffuse halo, therefore those targets were conservatively discarded from the final statistical analysis;
\item three candidate RHs, i.e. clusters hosting diffuse radio emission that spatially correlates with a highly disturbed X--ray morphology. The two most interesting cases are shown in Figure~\ref{fig:candidates}: residual, $\sim 800$~kpc--wide radio emission is still visible in the PSZ1G\,018.75+23.57 cluster after removing two point sources identified in the Northern VLA Sky Survey (NVSS, \cite{condon98}). As no spectral index information is available for these sources, they were subtracted by assuming a fiducial spectral index $\alpha = 0.7$. The errors introduced by the subtraction are $\sim 20\%$ of the thermal noise error estimates, as the proximity between the two observing frequencies makes spectral index uncertainties negligible. Including uncertainties in source subtraction, the flux density of the residual diffuse emission at 1.86~GHz is $S_{1.86} = 48.3 \pm 3.1$~mJy.

The second case in Figure~\ref{fig:candidates} referes to the Triangulum Australis cluster (\cite{scaife15}), which shows $\sim$Mpc scale diffuse radio emission offset, but clearly spatially correlated, with the X--ray brightness distribution. The comparison between higher angular resolution observations from the 843~MHz Sydney University Molonglo Sky Survey (SUMSS, \cite{bock99}) and the ESO Digitized Sky Survey optical image of the field reveals a spatial correspondence between some of the patchy 843~MHz radio emission and bright optical galaxies (see Figure~6 in \cite{bernardi16}). Although the point source contamination to the RH flux density measured at 1.86~GHz may not be negligible, its accurate estimate is not straightforward because of the uncertainty in the frequency extrapolation from 843~MHz and the still somewhat limited angular resolution of SUMSS, requiring further, dedicated observations; 
\item seven clusters for which no radio emission is detected down to the $0.3 - 0.8$~mJy~beam$^{-1}$ sensitivity limit and for which $P_{\rm 1.4} < 10^{24.4}$~Watt~Hz$^{-1}$ upper limits to the RH power could be set.
\end{itemize}
Upper limits to the RH power provide the most stringent test of the $P_{1.4} - M_{500}$ correlation and are plotted in Figure~\ref{fig:meerkat_correlation} together with literature measurements and the best estimates of the RH power for the PSZ1G\,018.75+23.57 and Triangulum Australis cases. Although the KAT--7 upper limits are compatible with the current correlation at the 95\% confidence level and, therefore, do not provide evidence for the presence of bimodality at low masses, they still show that bright RHs are rare in small clusters.

\section{MeerKAT's unique contribution}
\label{sec:MeerKAT_opportunities}

We are now in the position to show that the upcoming MeerKAT radio telescope is optimally placed to perfom crucial tests of current models of radio emission in clusters. With its 64, pseudo--randomly distributed antennas with a very dense 1~km core, MeerKAT offers a unique combination of excellent $uv$--coverage that is crucial in order to sample all the spatial scales spanned by RHs and the $\sim 5$~arcsec angular resolution necessary to identify and subtract discrete radio sources that may contaminate the RH emission. Moreover, with a 22~K system temperature\footnote{http://public.ska.ac.za/meerkat/meerkat-schedule} at 1.4~GHz, MeerKAT's pointed sensitivity is unmatched in the Southern Hemisphere.

\begin{figure*}
\centering
\includegraphics[width=1.2\columnwidth]{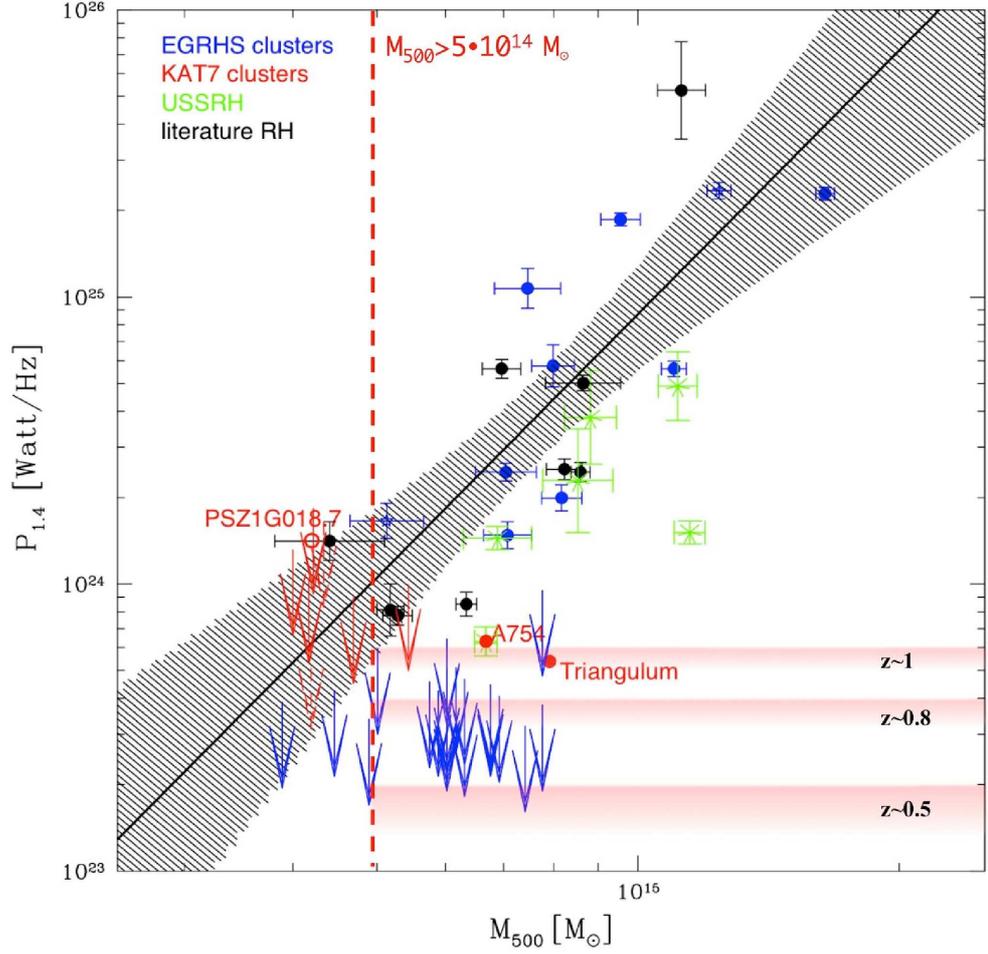}
\caption{Distribution of clusters in the $P_{\rm 1.4}-M_{500}$ diagram (adapted from \cite{bernardi16}). Different symbols indicate RHs belonging to the EGRHS (blue dots), RHs from the literature (black dots), RHs with very steep spectra (USSRH, green asterisks); upper limits from the EGRHS (blue arrows) and KAT--7 upper limits (red arrows). The best-fit relation to giant RHs only (black line) is shown together with its 95\% confidence region (shadowed region). The plot also includes the parameter space constrained by a possible MeerKAT survey of high redshift clusters (see text for details). The red horizontal bars indicate the lowest RH power detectable at the corresponding redshift for systems with $M_{500} > 5 \times 10^{14}$~M$_{\odot}$.}
\label{fig:meerkat_correlation}
\end{figure*}

MeerKAT is likely to be the only instrument in the Southern Hemisphere able to perform two critical tests:
\begin{itemize}
\item follow up observations of the KAT--7 sample. Given the current MeerKAT specifications, a 10 fold improvement in brightness sensitivity over the KAT--7 survey can be achieved in 10~hours per target, while avoiding the confusion limit. Such a survey will be able to either detect RHs or place upper limits that will be one order of magnitude below the current ones, i.e. well below the 95\% confidence level of the correlation. Given its angular resolution, it will also be able to identify, model and subtract point--like emission in the seven clusters where our current observations were not conclusive. Such a survey will provide the first statistically complete test of the current RH model in small systems, assessing the presence/absence of a bimodal distribution;
\item observations of high redshift systems. Very little is known about RHs in high redshift clusters, as statistically complete samples are observed only up to $z \sim 0.3$ (\cite{venturi08}, \cite{kale15}). Due to its very good angular resolution and $uv$ coverage, MeerKAT is again very well suited to observe higher redshift clusters whose RHs are naturally expected to be smaller. 
We selected 55 clusters from the South Pole Telescope SZ catalogue (\cite{reichardt13}) with $M_{500} > 5 \times 10^{14}$~M$_{\odot}$ ($80\%$ complete at $z > 0.4$). Assuming that the current $P_{1.4} - M_{500}$ correlation holds at higher redshift, Figure~\ref{fig:meerkat_correlation} shows the parameter space that this possible MeerKAT survey would constrain. Assuming a 5~hour observation per individual target, the survey would detect all the clusters at $z > 1$ with $P_{\rm 1.4} \gtrsim 6 \times 10^{23}$~Watt~Hz$^{-1}$, $z > 0.8$ with $P_{\rm 1.4} \gtrsim 4 \times 10^{23}$~Watt~Hz$^{-1}$ and $z > 0.5$ with $P_{\rm 1.4} \gtrsim 2 \times 10^{23}$~Watt~Hz$^{-1}$ respectively. Such a survey will complement the EGRHS (which provides the deepest and largest sample available to date) up to $z \sim 0.8$ at the same sensitivity level. Such a survey which will not only provide a critical test of current models, but also provide an independent probe of the merging history of clusters (\cite{cassano16}).
\end{itemize}

\section*{Acknowledgments}
This work is based on research supported in part by the National Research Foundation of South Africa (Grant Numbers 103424) and by the National Research Foundation under grant 92725. Any opinion, finding and conclusion or recommendation expressed in this material is that of the author(s) and the NRF does not accept any liability in this regard. This work was also partly supported by the Executive Programme of Scientific and Technological Co-operation between the Italian Republic and the Republic of South Africa 2014--2016 and by the National Research Foundation of South Africa (Grant Numbers 103424). The KAT--7 is supported by SKA South Africa and by the National Science Foundation of South Africa.

\end{document}